\documentclass[12pt]{article}
\usepackage{a4}

\let\oldbf\bf
\newcommand{\mbf}[1]{\ifmmode%
\mathchoice{\mbox{\boldmath{$\displaystyle#1$}}}%
{\mbox{\boldmath{$\textstyle#1$}}}%
{\mbox{\boldmath{$\scriptstyle#1$}}}%
{\mbox{\boldmath{$\scriptscriptstyle#1$}}}%
\else\oldbf#1\fi}

\begin{document}

\title{\bf Coherent transport}
\author{C. Henkel\\
\it Institute of Physics, Potsdam University,\\ 
\it Am Neuen Palais 10, 14469 Potsdam, Germany}
\date{29 November 2000; v2: 18 April 2001}

\maketitle

\vspace*{-70mm}
\noindent
Preprint physics/0012008,\\
accepted for publication in \textit{C. R. Acad. Sci. Paris, S\'erie IV} 
(2001)

\vspace*{70mm}

\begin{abstract}
We discuss the transport of matter waves in low-dimensional waveguides.
Due to scattering from uncontrollable noise fields, the spatial coherence 
gets reduced and eventually lost. We develop a description of this decoherence
process in terms of transport equations for the atomic Wigner function. 
We outline its derivation and discuss the special case of white noise
where an analytical solution can be found.
\end{abstract}

\section*{Introduction}

We discuss in this contribution the transport of atomic matter 
waves in a low-dimensional waveguide. Such structures may be created
close to solid substrates using electro-magnetic fields: the magnetic
field of a current-carrying wire combined with a homogeneous bias field,
e.g., gives rise to a linear waveguide \cite{Schmiedmayer98a,Anderson99,%
Prentiss00}. Planar waveguides may be constructed with repulsive magnetic
\cite{Hinds98} or optical \cite{Mlynek98b} fields that `coat' the substrate
surface. The atomic motion is characterised by bound vibrations in the
`transverse' direction(s) and an essentially free motion in the `longitudinal'
direction(s) along the waveguide axis (plane), respectively. Although direct
contact with the substrate is avoided by the shielding potential, the atoms
feel its presence through enhanced electromagnetic field fluctuations that
`leak' out of the thermal solid, typically held at room temperature. We have
shown elsewhere that these thermal near fields are characterised by a
fluctuation spectrum exceeding by orders of magnitude the usual blackbody
radiation \cite{Henkel99b,Henkel99c,Henkel00b,Henkel00c}. The scattering of
the atoms off the near field fluctuations occurs at a rate that may 
be calculated using Fermi's
Golden Rule. The consequences of multiple scattering is conveniently
described by a transport equation that combines in a self-consistent way
both ballistic motion and scattering. 

The purpose of this contribution is to outline a derivation of this
transport equation. The status of this equation is similar to that of
the quantum-optical master equations allowing to describe the evolution
of the reduced density matrix of an atomic system, on a time scale large
compared to the correlation time of the reservoir the system is coupled
to, typically the vacuum radiation field. In the case of transport
in waveguides, we 
face both temporal and spatial dynamics and therefore restrict 
our attention
to scales large compared to the correlation time and length of a
fluctuating noise potential. Our analysis uses a multiple scale
expansion adapted from \cite{Keller96}. Similar to the quantum-optical
case, we make an expansion in the perturbing potential to second order.
In the resulting transport equation, the noise is thus characterised
by its second-order correlation functions or, equivalently, its
spectral density. In the case of white noise, the transport equation 
can be explicitly solved. We have shown elsewhere \cite{Henkel00b} that
this approximation holds quite well for thermal near field fluctuations.
For technical noise, it also holds when the noise spectrum is flat on
a frequency scale roughly set by the `longitudinal' temperature of the 
atoms in the waveguide. The explicit solution yields an estimate for the
spatial coherence of the guided matter waves as a function of time. The
paper concludes with some remarks on the limits of validity of the present
transport theory. It cannot describe, e.g., Anderson localisation in one 
dimension \cite{Anderson58} because on the coarser spatial scale of the
transport equation, the scattering from the noise field is assumed to 
take place locally; interferences between different scattering sequences
are not taken into account. Decoherence in `curved' or `split' waveguides
also needs a refined theory because of the cross-coupling between the
transverse and longitudinal degrees of freedom, the former being `frozen
out' in our framework.

\section{Statistical matter wave optics}

The simplest model for atom transport in a low-dimensional waveguide
is based on the Schr\"odinger equation
\begin{equation}
{\rm i} \hbar \partial_t \psi( {\bf x}, t ) =
- \frac{ \hbar^2 }{ 2 m } \nabla^2 \psi 
+ V( {\bf x}, t ) \psi 
\label{eq:Schroedinger}
\end{equation}
The coordinate ${\bf x}$ describes the motion in the free waveguide 
directions. The transverse motion is `frozen out' by
assuming that the atom is cooled to the transverse ground state.
Atom-atom interactions are neglected, too.
$V( {\bf x}, t )$ is the noise potential: for a magnetic waveguide,
e.g., it is given by
\begin{equation}
V( {\bf {\bf x}}, t ) =  \langle s | \mbox{\boldmath$\mu$} \cdot 
{\bf B}( {\bf {\bf x}}, t ) | s \rangle 
,
\label{eq:perturbation}
\end{equation}
where $| s \rangle$ is the trapped internal state of the atom (we neglect
spin-changing processes), and ${\bf B}( {\bf x}, t )$ is the thermal
magnetic field. The noise potential is a statistical quantity
with zero mean and second-order correlation function
\begin{equation}
C_V( {\bf s}, \tau ) =
\langle 
V( {\bf x} + {\bf s}, t + \tau) \,
V( {\bf x}, t ) 
\rangle
,
\label{eq:potential-correlation}
\end{equation}
where the average is taken over the realisations of the noise potential.
We assume a statistically homogeneous noise, the correlation function
being independent of ${\bf x}$ and $t$. As a function of the separation
${\bf s}$, thermal magnetic fields are correlated on a length scale
$l_c$ given approximately by the distance $d$ between the waveguide axis 
and the
solid substrate \cite{Henkel00b}. This estimate is valid as long as the
wavelength $2\pi c/\omega$ corresponding to the noise frequency $\omega$ 
is large compared to $d$: for micrometre-sized waveguide structures,
this means frequencies below the optical range. 
The relevant frequencies of the noise will be
identified below and turn out to be much smaller than this.

The coherence properties of the guided matter waves are characterised
by the noise-averaged coherence function (the time dependence is suppressed
for clarity)
\begin{equation}
\rho( {\bf x}; {\bf s} )
=
\langle
\psi^*( {\bf x} - {\textstyle\frac12} {\bf s} )
\,
\psi( {\bf x} + {\textstyle\frac12} {\bf s} )
\rangle
.
\end{equation}
In complete analogy to quantum-optical master equations, this coherence
function may be regarded as the reduced density matrix of the atomic
ensemble, when the degrees of freedom of the noise are traced over.
The Wigner function gives a convenient representation of the coherence
function:
\begin{equation}
W( {\bf x}, {\bf p} ) = \int \!
\frac{ d^Ds }{ (2 \pi\hbar)^D } {\rm e}^{ - {\rm i} {\bf p} 
\cdot {\bf s} / \hbar }
\rho( {\bf x}; {\bf s} )
,
\label{eq:def-Wigner}
\end{equation}
where $D$ is the waveguide dimension. 
This representation allows to make a link to classical kinetic theory: 
$W( {\bf x}, {\bf p} )$ may be viewed as a quasi-probability in phase space.
For example, the spatial density $n({\bf x})$
and the current density ${\bf j}({\bf x})$
of the atoms are given by
\begin{eqnarray}
n({\bf x}) & = &
\int\!{\rm d}^Dp \, W( {\bf x}, {\bf p} )
\\
{\bf j}({\bf x}) & = &
\int\!{\rm d}^Dp \, \frac{ {\bf p} }{ m } W( {\bf x}, {\bf p} )
\end{eqnarray}
We also obtain information about the spatial coherence: the
spatially averaged coherence function $\Gamma( {\bf s}, t )$, for example, 
is related to the Wigner function by
\begin{eqnarray}
\Gamma( {\bf s}, t ) & \equiv &
\int{\rm d}^D x \, \rho( {\bf x}; {\bf s}, t )
\label{eq:def-Gamma}
\\
&=&
\int{\rm d}^Dx \,{\rm d}^Dp \,
{\rm e}^{ {\rm i} {\bf p} \cdot {\bf s} / \hbar } 
 \, W( {\bf x}, {\bf p}, t )
\end{eqnarray}
In the next section, we outline a derivation of a closed equation for
the Wigner function in terms of the noise correlation function.

\section{Transport equation}

Details of the derivation of the transport equation may be found in
the appendix~\ref{a:multiscale}. We quote here only the main assumptions
underlying the theory. 

\begin{itemize}
\item[(i)]
The noise potential is supposed to be \emph{weak} so that a perturbative
analysis is possible. As in quantum-optical master equations, a closed
equation is found when the expansion is pushed to second order in the
perturbation. 
\item[(ii)] The scale $l_c$ over which the noise is spatially correlated
is assumed to be small compared to the characteristic scale of variation
of the Wigner function. This implies a separation of the dynamics on
short and large spatial scales, the dynamics on the large scale 
being `enslaved' by certain averages over the short scale. 
Similarly, we assume that the potential fluctuates rapidly on the time scale
for the evolution of the Wigner function. 
These assumptions correspond to the Markov approximation of
quantum optics, where the master equation is valid on a coarse-grained time
scale. 
\end{itemize}

The derivation of the master equation is based on a multiple scale expansion.
Functions $f( {\bf x} )$ of the spatial coordinate are thus written in the
form 
\begin{equation}
f( {\bf x} ) = f( {\bf X}, \mbf{\xi} )
\end{equation}
where ${\bf X}$ gives the `slow' variation and the dimensionless variable
$\mbf{\xi} = {\bf x} / l_c$ gives the `rapid' variation on the scale of the
noise correlation length $l_c$. Spatial gradients are thus expanded using
\begin{equation}
\nabla_{\bf x} = \nabla_{\bf X} + \frac{1}{l_c} \nabla_{\mbf{\xi}}
\label{eq:gradient-expansion}
\end{equation}
By construction, the first term is much smaller than the second one. 
Finally, the Wigner function is expanded as
\begin{equation}
W( {\bf x}, {\bf p}, t ) =
W_0( {\bf X}, {\bf p}, t ) 
+
\eta^{1/2}
W_1( {\bf X}, \mbf{\xi}, {\bf p}, t ) 
+
{\cal O}( \eta )
\label{eq:multiple-scale-expansion}
\end{equation}
where $\eta \ll 1$ is the ratio between the correlation length $l_c$ and
a `macroscopic' scale on which the coordinate ${\bf X}$ varies. The
expansion allows to prove self-consistently that the zeroth order 
approximation $W_0$ does not depend on the short scale $\mbf{\xi}$,
and to fix the exponent $1/2$ for the first order correction.

The resulting transport equation specifies the evolution of the Wigner 
function $W_0$. Dropping the subscript $0$, it reads
\begin{eqnarray}
&&
\Bigl(
\partial_t + 
\frac{\bf p}m \cdot \nabla_{\bf x} 
\Bigr)
W( {\bf x}, {\bf p} )
=
\label{eq:transport-eq}
\\
&& 
\int\!{\rm d}^Dp' \,
S_V( {\bf p}' - {\bf p}, E_{{\bf p}'} - E_{\bf p} )
\left[
W( {\bf x}, {\bf p}' )
-
W( {\bf x}, {\bf p} )
\right]
,
\nonumber
\end{eqnarray}
where $S_V$, the \emph{spectral density} of the noise, is essentially
the spatial and time Fourier transform of the noise correlation function
\begin{eqnarray}
&&
S_V( {\bf q}, \Delta E) =
\frac{ 1 }{ \hbar^2 }
\int\!\frac{ {\rm d}^D s \, {\rm d}\tau}{ (2\pi\hbar)^{D} }
C_V( {\bf s}, \tau )
\, {\rm e}^{ - i ( {\bf q}\cdot{\bf s} - \Delta E \tau ) / \hbar}
.
\label{eq:potential-spectrum}
\end{eqnarray}
The left hand side of the transport equation gives the free ballistic
motion of the atoms in the waveguide. If an external force were
applied, an additional term ${\bf F} \cdot \nabla_{\bf p}$ would
appear. The right hand side describes the scattering from the noise
potential. $E_{\bf p} = p^2 / 2m$ is the de Broglie dispersion
relation for matter waves. We observe that scattering processes 
${\bf p} \to {\bf p}'$ occur at a rate given by the noise
spectrum at the Bohr frequency $(E_{\bf p} - E_{{\bf p}'})/\hbar$.
If the potential noise is static (as would be the
case for a `rough potential'), then its spectral density is
proportional to $\delta( \Delta E )$, and energy is conserved.
If we are interested in the scattering between guided momentum states,
then the initial and final energies $E_{\bf p}$, $E_{{\bf p}'}$ are typically
of the order of the (longitudinal) temperature $kT$ of the ensemble.
The relevant frequencies in the noise spectral density are thus
comparable to $kT/\hbar$.

\section{Results}

\subsection{White noise}

White noise is characterised by a constant spectral density, i.e., the
noise spectrum $S_V( {\bf q}, \Delta E )$ is independent of $\Delta E$.
Equivalently, the noise correlation is $\delta$-correlated in time:
\begin{equation}
C_V( {\bf s}, \tau ) = B_V( {\bf s} )\, \delta( \tau )
.
\end{equation}
The integration over the momentum ${\bf p}'$ in~(\ref{eq:transport-eq})
is now not restricted by
energy conservation, and the right hand side of the transport 
equation becomes a convolution. One therefore
obtains a simple solution using Fourier transforms. Denoting ${\bf k}$
(dimension: wavevector) and ${\bf s}$ (dim.: length) the Fourier variables 
conjugate to ${\bf x}$ and ${\bf p}$, we find the equation
\begin{eqnarray}
\Bigl(
\partial_t + 
\frac{ \hbar {\bf k}}{ m } \cdot \nabla_{\bf s} 
\Bigr)
\tilde W( {\bf k}, {\bf s} )
=
- \gamma( {\bf s} ) 
\tilde W( {\bf k}, {\bf s} )
.
\label{eq:Fourier-transport}
\end{eqnarray}
where we have introduced the rate
\begin{equation} 
\gamma( {\bf s} ) 
= \frac{ 1 }{ \hbar^2 } 
\left(
B_V( {\bf 0} ) - B_V( {\bf s} )
\right)
.
\end{equation}
Eq.(\ref{eq:Fourier-transport}) is easily solved using the method of 
characteristics, using ${\bf s} - \hbar{\bf k}t/m$ as a new variable. 
One finds
\begin{eqnarray}
\tilde{W}( {\bf k}, {\bf s}; t ) & = &
\tilde{W}_i( {\bf k}, {\bf s} - \hbar{\bf k}t/m ) 
\times {}
\nonumber\\
&& {} \times  
\exp{\left[
- 
\int_0^t\!{\rm d}t'
\gamma( {\bf s} - \hbar{\bf k} t' /m )
\right]
}
,
\label{eq:analytic-solution}
\end{eqnarray}
where $\tilde{W}_i( {\bf k}, {\bf s} )$ is the Wigner function at $t = 0$.

We observe in particular that the spatially averaged coherence 
function~(\ref{eq:def-Gamma}) shows an exponential decay as time
increases:
\begin{equation}
{\Gamma}( {\bf s}; t ) = {\Gamma}_i( {\bf s} )
\exp{\Bigl[ 
- 
\gamma( {\bf s} ) t
\Bigr]}
.
\label{eq:analytic-solution-2}
\end{equation}
We can thus give a physical meaning to the quantity $\gamma( {\bf s} )$:
it is the rate at which two points in the matter wave field, that are
separated by a distance ${\bf s}$, lose their mutual coherence. This
rate saturates to $\gamma = \gamma( \infty ) = B_V( {\bf 0} ) / \hbar^2$ for
distances $s \gg l_c$ large compared to the correlation length of the
noise field (the correlation $B_V( {\bf s} )$ then vanishes).
This saturation has been discussed, e.g., in~\cite{Cheng99}. 
As shown in~\cite{Henkel00c}, the rate $\gamma$ is equal to the
total scattering rate from the noise potential, as obtained from Fermi's
Golden Rule. For distances smaller than $l_c$, the decoherence rate
$\gamma( {\bf s} )$ decreases since the two points of the matter wave
field `see' essentially the same noise potential. The exact 
solution~(\ref{eq:analytic-solution-2}) thus implies that after a time
of the order of the scattering time $1/\gamma$, the spatial coherence
of the atomic ensemble has been reduced to the correlation length $l_c$.
The estimates given in~\cite{Henkel00c} imply a time scale of the
order of a fraction of a second for waveguides at a micrometre distance
from a (bulk) metallic substrate. Significant improvements can be made using
thin metallic layers or wires, nonconducting materials or by mounting 
the waveguide at a larger distance from the substrate~\cite{Henkel00c}.

At timescales longer than the scattering time $1/\gamma$, the spatial
coherence length of the atoms decreases more slowly, approximately as
$l_c/ \sqrt{ \gamma t }$~\cite{Henkel00c}. This is due to a diffusive
increase of the width of the atomic momentum distribution, with a 
diffusion constant of the order of $D = \hbar^2 \gamma / l_c^2$. This
constant is in agreement with a random walk in momentum space: for each
scattering time $1/\gamma$, the atoms absorb a momentum $q_c = \hbar/l_c$ 
from the noise potential. The momentum step $q_c$ follows from the fact
that the noise potential is smooth on scales smaller than $l_c$, its Fourier 
transform therefore contains momenta up to $\hbar/l_c$.

\subsection{Fokker-Planck equation}

The momentum diffusion estimate given above can also be retrieved from
the transport equation, making an expansion of the Wigner distribution
as a function of momentum. We assume that the typical momentum transfer
$q_c$ absorbed from the noise is small compared to the scale of variation
of the Wigner distribution, and expand the latter to second order.
This manipulation casts the transport equation into a Fokker-Planck
form
\begin{eqnarray}
&&
\Bigl(
\partial_t 
+ 
\frac{\bf p}m \cdot \nabla_{\bf x} 
+
{\bf F}_{\rm dr}( {\bf p} ) \cdot \nabla_{\bf p}
\Bigr)
W( {\bf x}, {\bf p} )
=
\label{eq:Fokker-Planck}
\\
&& \qquad
\sum_{ij}
D_{ij}( {\bf p } )
\frac{ \partial^2 }{ \partial p_i \partial p_j }
W( {\bf x}, {\bf p} )
,
\nonumber
\end{eqnarray}
where the drift force and the diffusion coefficient are given by
\begin{eqnarray}
{\bf F}_{\rm dr}( {\bf p} ) &=&
- \int\!{\rm d}^D q \,
{\bf q} \,
S_V( {\bf q}, E_{\bf p + q} - E_{\bf p} )
\label{eq:drift}
\\
D_{ij}( {\bf p } )
&=&
\int\!{\rm d}^D q \,
q_i q_j \,
S_V( {\bf q}, E_{\bf p + q} - E_{\bf p} )
.
\label{eq:diffusion}
\end{eqnarray}
In the special case of white noise, the ${\bf p}$-dependence of these
quantities drops out. Also the drift force is then zero because the
noise correlation function is real and the spectrum $S_V( {\bf q} )$
even in ${\bf q}$. Since $q_c$ gives the width of the spectrum,
the diffusion coefficient turns out to be of order $q_c^2 \gamma$,
as estimated before. 

Casting the transport equation into Fokker-Planck form, one can easily
take into account the scattering from the noise field in (classical)
Monte Carlo simulations of the atomic motion: one simply has to add
a random force whose correlation is given by the diffusion coefficient.

We note, however, that the Fokker-Planck equation cannot capture the
initial stage of the decoherence process, starting from a wave field
that is coherent over distances larger than the correlation length $l_c$.
Indeed, it may be shown (neglecting the ${\bf p}\cdot\nabla_{\bf x}$ term
and the drift force, assuming an isotropic diffusion tensor for simplicity) 
that~(\ref{eq:Fokker-Planck}) yields a spatially averaged coherence function 
\begin{equation}
\Gamma_{FP}( {\bf s}, t ) =
\Gamma_i( {\bf s} ) \,
\exp\left[ - 
D s^2 t / \hbar^2 \right]
\label{eq:decoherence-FP}
\end{equation}
This result implies a decoherence rate proportional to $s^2$ without
saturation. It is hence valid only at large times (compared to the
scattering time $1/\gamma$) where the exponentials in 
both solutions~(\ref{eq:analytic-solution-2}, \ref{eq:decoherence-FP})
are essentially zero for $s \ge l_c$.

\section{Concluding remarks}

We have given an outline of a transport theory for dilute atomic gases
trapped in low-dimensional waveguides. This theory allows to follow the
evolution of the atomic phase-space distribution (more precisely,
the atomic, noise-averaged Wigner function) when the atoms are subject
to a noise potential with fluctuations in space and time. The spatial
coherence of the gas can be tracked over temporal and spatial
scales larger than the correlation scale of the noise, in a manner 
similar to the master equations of quantum optics. We have given
explicit results in the case of white noise, highlighting spatial
decoherence and momentum diffusion. 

The transport equation has to be taken with care for strong noise
potentials because its derivation is based on second-order perturbation
theory. It is certainly not valid when the `mean free path' $\sim
\bar{v} / \gamma$ ($\bar v$ is a typical velocity of the gas) is
smaller than the noise correlation length $l_c$ because then the
Wigner distribution changes significantly over a small spatial
scale. (In technical terms, the approximation of a local scattering kernel 
in~(\ref{eq:transport-eq}) is no longer appropriate.) Also, the 
theory cannot describe Anderson localisation in 1D waveguides
with static noise~\cite{Anderson58}. This can be seen by working
out the scattering kernel with $S_V( {q}, \Delta E ) =
S_V( {q} )\, \delta( \Delta E )$:
\begin{eqnarray}
&&
2m \int\!{\rm d}p' \,
S_V( p' - p )\, \delta( p'^2 - p^2 ) 
\left[
W( x, p' ) - W( x, p )
\right]
\nonumber
\\
&&
=
\frac{ m S_V( 2 p ) }{ p }
\left[
W( x, -p ) - W( x, p )
\right]
.
\end{eqnarray}
We find a divergence of the scattering rate at $p \to 0$
since the spectrum $S_V( 2 p )$ is finite in this limit. The 
one-dimensional, static case therefore merits further investigation.
We also mention that is has been found recently that
Anderson localisation is destroyed when time-dependent fluctuations 
are superimposed on the static disorder~\cite{Flores99,Gurvitz00}. 
In this context, transport (or master) equations similar to our approach 
have been used.

\paragraph{Acknowledgements.}
We thank S. A. Gardiner, S. P\"otting, M. Wilkens, and P. Zoller 
for constructive discussions. Continuous support from M. Wilkens is
gratefully acknowledged.

\appendix
\section{Multiple scale derivation of the transport equation}
\label{a:multiscale}

The Schr\"odinger equation~(\ref{eq:Schroedinger}) gives the following
equation for the Wigner function
\begin{eqnarray}
&&\left(
\partial_t + {\bf p}\cdot\nabla_{\bf x} 
\right)
W( {\bf x}, {\bf p} ) =
\label{eq:for-Wigner-1}\\
&&
- \frac{ {\rm i} }{ \hbar }
\int \!\frac{ {\rm d}^D q }{ (2\pi\hbar)^D }
\tilde{V}( {\bf q}, t )
\,{\rm e}^{ {\rm i} {\bf q} \cdot {\bf x} }
\left[ 
W( {\bf x}, {\bf p} + {\textstyle\frac12} {\bf q} )
-
W( {\bf x}, {\bf p} - {\textstyle\frac12} {\bf q} )
\right]
\nonumber
\end{eqnarray}
where $\tilde{V}( {\bf q}, t )$ is the spatial Fourier transform of the
noise potential. Since this potential is assumed weak and varies on
a scale given by the correlation length $l_c$, we introduce the
following scaling
\begin{equation}
\tilde{V}( q_c {\bf u}, t ) = 
\int\!{\rm d}^Dx \,
{\rm e}^{ - {\rm i} q_c {\bf u} \cdot {\bf x} / \hbar }
V( {\bf x}, t )
=
l_c^D \eta^\beta \hat V( {\bf u}, t )
\label{eq:scaling}
\end{equation}
where $q_c \equiv \hbar / l_c$ is the typical momentum width of 
$\tilde{V}( {\bf q}, t )$ and ${\bf u}$ is a dimensionless vector.
The parameter $\eta$ is given by the ratio between the small
scale $l_c$ and the `macroscopic' scale of the position distribution,
the (positive) exponent $\beta$ remains to be determined.
We assume $\eta \ll 1$ and make the multiple scale 
expansion~(\ref{eq:multiple-scale-expansion}) for the Wigner function.
Using the expansion~(\ref{eq:gradient-expansion}) for the spatial
gradient, we get
\begin{eqnarray}
&&
\left[
\partial_t + \frac{\bf p}m \cdot \left(
\nabla_{\bf X} + \frac{1}{l_c} \nabla_{\mbf{\xi}}
\right)
\right]
\left(
W_0 + \eta^\alpha W_1 
\right)
=
\label{eq:for-I-1}
\\
&&- 
\frac{ {\rm i} \eta^\beta }{ q l_c }
\int\!\frac{ {\rm d}^D u }{ (2\pi)^D }
\hat V( {\bf u}, t )
\,{\rm e}^{ {\rm i} {\bf u} \cdot {\bf x} / l_c }
\left[
W( {\bf x}, {\bf p} + q_c {\bf u}/2 )
-
W( {\bf x}, {\bf p} - q_c {\bf u}/2 )
\right]
\nonumber
\end{eqnarray}
We now take the limit $\eta \to 0, \, l_c \to 0$ at fixed $q_c$.
The most divergent term on the left hand side is the one with
$(1/l_c) \nabla_{\mbf{\xi}}W_0$. It could only be balanced with
a term on the right hand side involving $W_0$, but due to the
small factor $\eta^\beta$, this term cannot have the same order of
magnitude. We must therefore require that $(1/l_c) \nabla_{\mbf{\xi}}W_0$
vanishes individually: the zeroth order Wigner function is 
independent of the short scale variable $\mbf{\xi}$.

The next terms on the left hand side contain $(\eta^\alpha/l_c) 
\nabla_{\mbf{\xi}}W_1$ and $\nabla_{\bf X} W_0$, while on the
right hand side the leading order is $(\eta^\beta/l_c) W_0$.
We look for a connection between $W_0$ and $W_1$, and 
therefore, the left hand $W_1$ term must be more divergent than the 
$W_0$ term. 
This is the case if $\eta^\alpha {\cal O}( 1 / l_c ) 
\gg {\cal O}( 1 / X ) \sim \eta {\cal O}( 1 / l_c )$. 
We thus conclude that $\alpha < 1$. Comparing powers of $\eta$
on the left and right hand side, we find $\alpha = \beta$, 
since the vector $\bf u$ and the scaled distance $\mbf{\xi}$ are
of order unity. Therefore we get the equation
\begin{eqnarray}
&&
\left(
l_c \partial_t + \frac{\bf p}m \cdot \nabla_{\mbf{\xi}}
\right)
W_1( {\bf X}, \mbf{\xi}, {\bf p} )
=
\\
&& 
- \frac{ {\rm i} }{ q_c }
\int\!\frac{ {\rm d}^D u }{ (2\pi)^D }
\hat V( {\bf u}, t )
\,{\rm e}^{ {\rm i} {\bf u} \cdot \mbf{\xi} }
\left[
W_0( {\bf X}, {\bf p} + q_c {\bf u}/2 )
-
W_0( {\bf X}, {\bf p} - q_c {\bf u}/2 )
\right]
\nonumber
\end{eqnarray}
In the exponential, only the short length scale 
$\mbf{\xi} = {\bf x} / l_c$ occurs.
We thus find that the large scale variable ${\bf X}$ is a parameter in this
equation, and get a solution via Fourier transforms with respect 
to $\mbf{\xi}$ and $t$. In the spirit of the Markov approximation, we
take the slowly varying $W_0$ (as a function of time) out of the time
integral
\begin{equation}
\int_{-\infty}^{\infty}\!{\rm d}t \,
{\rm e}^{ {\rm i} \omega t }
\hat V( {\bf u}, t ) \,
W_0( \ldots, t ) 
\approx
W_0( \ldots, t ) 
\hat V[ {\bf u}, \omega ] 
\end{equation}
where $\hat V[ {\bf u}, \omega ]$ denotes the double space and time Fourier
transform of the potential. We note $\mbf{\kappa}$, $\omega$ the conjugate 
variables for the spatial Fourier transform and find the following solution
for the first order Wigner function
\begin{eqnarray}
&&
W_1( {\bf X}, \mbf{\xi}, {\bf p} ) =
\label{eq:result-W1}
\\
&&
- \frac{ \rm i }{ q_c }
\int\!\frac{ {\rm d}\omega }{ 2\pi }
\int\!\frac{ {\rm d}^D\kappa }{ (2\pi)^D }
\frac{ {\rm e}^{ {\rm i} \mbf{\kappa} \cdot \mbf{\xi} - {\rm i} \omega t}
\hat V[ \mbf{\kappa}, \omega ] }{
{\rm i} \mbf{\kappa} \cdot {\bf p}/m 
- {\rm i} l_c \omega + 0 }
\left(
W_0( {\bf X}, {\bf p} + q_c \mbf{\kappa}/2 )
-
W_0( {\bf X}, {\bf p} - q_c \mbf{\kappa}/2 )
\right)
\nonumber
\end{eqnarray}
The $+0$ prescription in the denominator is related to causality: it
ensures that the poles in the complex $\omega$-plane are moved into the lower 
half plane, avoiding a blow-up of $W_1$. 

This result will be inserted into the next order equation that also links
$W_0$ to $W_1$:
\begin{eqnarray*}
&& \left(
\partial_t +
\frac{\bf p}m \cdot \nabla_{\bf X} 
\right)
W_0 = 
\label{eq:W0-linked-to-W1}
\\
&&
- \frac{ {\rm i} \eta^{2\alpha} }{ q_c l_c }
\int\!\frac{ {\rm d}^Du }{ (2\pi)^D } 
\hat V( {\bf u}, t )
\, {\rm e}^{ {\rm i} {\bf u} \cdot \mbf{\xi} }
\left[
W_1( {\bf X}, \mbf{\xi}, {\bf p} + q_c{\bf u}/2 )
-
W_1( {\bf X}, \mbf{\xi}, {\bf u} - q_c{\bf u}/2 )
\right]
\end{eqnarray*}
Note that this equation is scaled consistently if ${\cal O}( 1 / X) \sim
\eta^{2\alpha} {\cal O}( 1 / l_c ) = \eta^{2\alpha - 1}
{\cal O}( 1 / X )$. This determines the exponent $\alpha = \frac12$.
The result is an equation for $W_0$ only. We take the statistical
average and make the factorisation
\begin{equation}
\langle
\hat V( {\bf u}, t ) 
\,
\hat V[ \mbf{\kappa}, \omega ]
\,
W_0( {\bf X}, {\bf p} )
\rangle
=
\langle
\hat V( {\bf u}, t ) 
\,
\hat V[ \mbf{\kappa}, \omega ]
\rangle
\,
W_0( {\bf X}, {\bf p} )
.
\end{equation}
This may be justified heuristically as follows: it seems reasonable
that the statistical average can also be performed via `spatial coarse
graining', \emph{i.e.}, taking an average over the small-scale fluctuations
of the medium. This is precisely the picture behind transport theory: 
the individual scattering events are not resolved but only the behaviour
of the matter wave on larger scales. 
The lowest order Wigner function $W_0$ may be taken out of 
the coarse grain average because it does not depend on the short
scale $\mbf{\xi}$ by construction.

Finally, we introduce the spectral density $\hat S( {\bf u}, \omega )$ 
of the (scaled) noise potential
\begin{equation}
\langle
\hat V( {\bf u}, t ) 
\,
\hat V[ \mbf{\kappa}, \omega ]
\rangle
= 
(2\pi)^D \hat S( {\bf u}, \omega ) \,
{\rm e}^{ {\rm i} \omega t } \,
\delta( {\bf u} + \mbf{\kappa} )
\end{equation}
This allows to perform the integration over $\mbf{\kappa}$ 
when~(\ref{eq:result-W1}) is inserted into~(\ref{eq:W0-linked-to-W1}).
The result still contains a frequency integral where denominators
of the following form appear
\begin{equation}
\frac{ 1 }{
{\rm i} ({\bf u}/ m )\cdot
\left(
{\bf p} + q_c {\bf u} / 2 
\right) - {\rm i} l_c \omega + 0
}
= \frac{ - {\rm i} q_c }{ 
E_{{\bf p}+q_c{\bf u}} - E_{\bf p} - \hbar\omega - {\rm i} 0 }
\label{eq:resonant-denominator}
\end{equation}
A second term contains the sign-reversed energy difference. These
denominators ensure that the kinetic energy change occurring in
the scattering is compensated by a `quantum' $\hbar\omega$ 
from the noise potential. 

We write the denominators~(\ref{eq:resonant-denominator})
as a $\delta$-function plus a principal part.
For the classical noise potential considered here, 
the power spectrum $\hat S( {\bf u}, \omega )$ is
even in $\omega$, so that the $\delta$-functions combine and 
the principal parts drop out. We finally get
\begin{eqnarray}
&& \left(
\partial_t +
\frac{\bf p}m \cdot \nabla_{\bf X} 
\right)
W_0 = 
\label{eq:W0-transport}
\\
&&
\frac{ \eta }{ \hbar^2 }
\int\!\frac{ {\rm d}^Du }{ (2\pi)^D } 
\hat S( {\bf u}, \Delta E / \hbar )
\left[
W_0( {\bf X}, {\bf p} + q_c{\bf u} )
-
W_0( {\bf X}, {\bf p} )
\right]
\nonumber
\end{eqnarray}
where $\Delta E = E_{{\bf p}+q_c{\bf u}} - E_{\bf p}$.
It is easily checked that this is the transport 
equation~(\ref{eq:transport-eq}), taking into account the relation
between the scaled and non-scaled noise spectra
\begin{equation}
\frac{ \eta l_c^3 }{ \hbar^2 } \hat S_V( {\bf u}, \Delta E / \hbar )
= S_V( q_c {\bf u}, \Delta E )
\end{equation}
that follows from~(\ref{eq:potential-spectrum}) and (\ref{eq:scaling}).


\begin{thebibliography}{10}

\bibitem{Schmiedmayer98a}
J. Schmiedmayer, Eur. Phys. J. D {\bf 4},  57  (1998).

\bibitem{Anderson99}
D. M\"uller, D.~Z. Anderson, R.~J. Grow, P.~D.~D. Schwindt, and E.~A. Cornell,
  Phys. Rev. Lett. {\bf 83},  5194  (1999).

\bibitem{Prentiss00}
N.~H. Dekker {\it et~al.}, Phys. Rev. Lett. {\bf 84},  1124  (2000).

\bibitem{Hinds98}
E.~A. Hinds, M.~G. Boshier, and I.~G. Hughes, Phys. Rev. Lett. {\bf 80},  645
  (1998).

\bibitem{Mlynek98b}
H. Gauck, M. Hartl, D. Schneble, H. Schnitzler, T. Pfau, and J. Mlynek, Phys.
  Rev. Lett. {\bf 81},  5298  (1998).

\bibitem{Henkel99b}
C. Henkel and M. Wilkens, Europhys. Lett. {\bf 47},  414  (1999).

\bibitem{Henkel99c}
C. Henkel, S. P{\"o}tting, and M. Wilkens, Appl. Phys. B {\bf 69},  379
  (1999).

\bibitem{Henkel00b}
C. Henkel, K. Joulain, R. Carminati, and J.-J. Greffet, Opt. Commun. {\bf 186}
  (2000) 57.

\bibitem{Henkel00c}
C. Henkel and S. P\"otting, Appl. Phys. B {\bf 72} (2001) 73 
  (selected papers of the Bonn 2000 DPG meeting).

\bibitem{Keller96}
L. Ryzhik, G. Papanicolaou, and J.~B. Keller, Wave Motion {\bf 24},  327
  (1996).

\bibitem{Anderson58}
P.~W. Anderson, Phys. Rev. {\bf 109},  1492  (1958).

\bibitem{Cheng99}
C.-C. Cheng and M.~G. Raymer, Phys. Rev. Lett. {\bf 82},  4807  (1999).

\bibitem{Flores99}
J.~C. Flores, Phys. Rev. B {\bf 60},  30  (1999).

\bibitem{Gurvitz00}
S.~A. Gurvitz, Phys. Rev. Lett. {\bf 85},  812  (2000).

\end{thebibliography}

\end{document}